\documentclass[twocolumn,prb,showpacs]{revtex4}

\usepackage{graphicx}

\begin{document}

\title{{\rm\sc Phys. Rev. B 68, 134445 (2003)}\\[.4cm]Luttinger liquid behavior in spin chains with a magnetic field}

\author{G\'abor F\'ath}
\affiliation{Research Institute for Solid State Physics and Optics, P.O. Box 49,
         H-1525 Budapest, Hungary}

\date{8 April 2003}

\begin{abstract}
Antiferromagnetic Heisenberg spin chains in a sufficiently strong magnetic
field are Luttinger liquids, whose parameters depend on the
actual magnetization of the chain. Here we present precise numerical estimates
of the Luttinger liquid dressed charge $Z$, which determines the critical exponents, by
calculating the magnetization and quadrupole operator profiles for $S=1/2$ and $S=1$
chains using the density matrix renormalization group method. Critical amplitudes
and the scattering length at the chain ends are also determined. Although both systems
are Luttinger liquids the characteristic parameters differ considerably.
\end{abstract}

\maketitle

\section{Introduction}

The one-dimensional (1D) antiferromagnetic Heisenberg chain
\begin{equation}
    H =\sum_{i=1}^{N-1} {\rm\bf S}_i  {\rm\bf S}_{i+1}-h \sum_{i=1}^N S^z_i
    \label{HAF}
\end{equation}
is one of the most thoroughly investigated paradigms of interacting many-body
systems. It is well-known that for zero magnetic field the low-energy physics
depends very much on the spin length $S$. As Haldane predicted in 1983,\cite{Hal-gap-83}
integer $S$ chains possess an energy gap $\Delta_g(S)$ above the ground state, the size
of which vanishes exponentially as $S\to\infty$, whereas half-integer $S$ chains are
gapless. The presence of the Haldane gap in integer spin chains implies the occurrence of
a critical field $h_c=\Delta_g$ beyond which the system gets magnetized. As the
gap collapses at the critical field the ground state structure changes adequately, and in
an already magnetized state there remains no conceptual difference in the low-energy properties
between integer and half-integer chains.

Partially magnetized antiferromagnetic Heisenberg chains at low energies are expected to be one-component
Luttinger liquids\cite{Hal-JPC81} (LLs) irrespective of the spin length $S$. This is known
rigorously in the $S=1/2$ case where the model can be analyzed using the Bethe Ansatz.\cite{Hal-PRL80} The
validity of the Luttinger liquid description for $S\ge 1$ chains and coupled spin-1/2 ladders has been
investigated and confirmed by analytical\cite{Tsvelik,Chi-Gia,Kon-Fen} and numerical\cite{Sak-Tak,Usa-Sug,Hik-Fur}
methods. Although there is a theoretical possibility of finding gapful behavior
(magnetization plateau) at special values of the magnetization, such a scenario does not
seem to be realized in pure nearest-neighbor Heisenberg chains (see Ref.\ \onlinecite{Osh-Yam-Aff}
for $S=3/2$). Neither would we expect multi-component LLs, which can otherwise occur
for more general (higher-order) couplings.\cite{Fat-Lit}

The LL concept bears a strong relationship to $c=1$ conformal field
theory (CFT).\cite{Gog-Ner-Tsv} In fact the LL is hardly more than the CFT adapted to the situation
with two Fermi points $2k_F$ momentum apart. The one-component LL is a three parameter
theory.\cite{Hal-JPC81}
In the present context the first parameter, the location of the Fermi points $\pm k_F$, is determined
by the magnetization through
\begin{equation}
   2k_F = 2\pi (S-m),
   \label{m_kF}
\end{equation}
where $m=S^z_{\rm tot}/L$ is the (bulk) magnetization.
The second parameter, the Fermi velocity $v$ is just an energy scale, whereas the third
parameter, to be called the "dressed charge" $Z$, determines the universality class and
the critical exponents. (In the bosonization literature the notation $K=Z^2$ is standard.)
The traditional LL parameters, the velocities for charge and
current excitations can be expressed with $Z$ as $v_{\rm charge}=v/Z^2$,
$v_{\rm current}=vZ^2$.\cite{Hal-PRL81}

All three parameters are functions of the actual magnetization $m$ and
the spin length $S$. 
The LL parameters of the $S=1$ chain in a magnetic field have already been determined to some extent
using numerical methods. Sakai and Takahashi diagonalized small finite chains up to $N=16$ with
periodic boundary condition and used the prediction of CFT on finite size energy spectra
to estimate the critical exponents.\cite{Sak-Tak} A similar method was used by Usami and Suga\cite{Usa-Sug}
for the $S=1/2$ ladder which behaves as a Haldane-gap system for strong ferromagnetic interchain interaction.  
More recently Campos Venuti {\em et al.} used the DMRG to compute directly
the transverse two-point function $G^{xx}(r)=\langle S^x(0) S^x(r) \rangle$ on chains with $N=80$ and
fitted these data to the expected asymptotic form of $G^{xx}$.\cite{Campos-Venuti}

In this paper we apply an alternative method. We calculate precise numerical estimates to the critical exponents
and by this to the LL dressed charge $Z$ via a direct numerical determination of the magnetization
profiles in finite open chains. The magnetization profile is, by definition, the positional
dependence of the local magnetization $m_n=\langle S^z_n \rangle$ in an open chain with some
well-defined boundary condition. The form of the profile depends on the applied
boundary condition, and may involve surface exponents as well whenever the boundary condition on the
left and right ends differ. In order to simplify the expected behavior we will only consider cases where
the boundary condition is identically open (free) on both ends; in this case only bulk exponents
come into play.\cite{BurkEise}
In a semi-infinite chain, far from the chain's end, the magnetization profile is expected to
decay to its bulk value $m$ algebraically as
\begin{equation}
   m(r) \simeq m + A_m \cos(2k_F r+\phi)\, r^{-\eta_z/2},
   \label{semiinf}
\end{equation}
where $A_m$ is a non-universal amplitude, $\phi$ is a phase shift, $\eta_z$ is the
(bulk) critical
exponent, defined through the translation invariant longitudinal two-point function
$G^{zz}(r)=\langle S^z(0) S^z(r)\rangle-m^2 \simeq A \cos(2k_F r+\phi)\, r^{-\eta_z}$,
and $k_F$ is determined as a function of $m$ by Eq.\ (\ref{m_kF}). The form of the {\em Friedel oscillation}
in Eq.\ (\ref{semiinf}) contains indirect information about the LL parameter $Z$, and this can be exploited in a
numerical procedure
to calculate precise estimates. This program has been carried out for the $S=1/2$ chain and ladders in Ref.\ 
\onlinecite{Hik-Fur}, for the Kondo lattice model in Ref.\ \onlinecite{Shibata}, for the Hubbard model in
Ref.\ \onlinecite{Bedurftig}, and for the $t-J$ model in Ref.\ \onlinecite{White-Affleck}.

While the algebraic decay with $\eta_z/2$ is a standard consequence of
criticality,\cite{Fis-deG} the $2k_F$
oscillation is a special LL feature, which stems from the fact that we work with two families
of (chiral) CFT operators living around the two Fermi points in $k$-space. Without
the $2k_F$ oscillation it is a standard exercise in CFT to derive the shape of the
magnetization profile in a strip geometry of width $L$ by applying the logarithmic
mapping.\cite{Cardy,BurkEise} This transformation, together with the proper account
for the $2k_F$ LL term yields a prediction for the magnetization
profile of a finite Luttinger liquid segment of length $L$ with open boundary condition on
both ends\cite{Hik-Fur,White-Affleck}
\begin{equation}
   m^{(L)}(r) \simeq m + A_m \cos\!\left[2k_F\!\! \left(r-\frac{L}{2}\right)\right]
   \left[ {L\over \pi} \sin\!\left(\pi\frac{r}{L}\right)\right]^{-\eta_z/2} \!\!\!.
   \label{profile1}
\end{equation}
Note that there is no need to introduce explicitly a phase shift $\phi$ as in the
semi-infinite case Eq.\ (\ref{semiinf}), since the symmetry
of the profile with respect to $r\to L-r$ implies that $\phi=-k_F L$.

The predicted LL profile in Eq.\ (\ref{profile1}) is based on the continuum limit.
However, for finite lattice systems such as the Heisenberg chain, we expect corrections.
Conformal invariance and thus Eq.\ (\ref{profile1}) is only expected to be valid
asymptotically in the large $L$ limit with $r$ satisfying $0<<r<<L$. Clearly, in a strict
sense the magnetization cannot diverge at the chain ends since $m_n\le S$.
Phenomenologically this natural cut-off at the boundary acts as an effective impurity
put into the CFT model at the system ends. This defines an effective "scattering length",
$\delta N={\cal O}(1)$, associated with the boundary, suggesting that we should replace the naive
system size $L=N$ in Eq.\ (\ref{profile1}) with an "effective" system size
$L\to N+2\,\delta N$. Based on this intuitive argument in the following we will work with a slightly
modified Ansatz 
\begin{eqnarray}
   m^{(N)}_n \!&\simeq&\! m + A_m \cos\left[2k_F \left(n-{N+1\over 2}\right)\right] \nonumber\\ 
       &&\times \left[ {N+2\,\delta N\over \pi} \sin\!\left(\pi\frac{n-1/2+\delta N}{N+2\,\delta N}\right)\right]^{-\eta_z/2}
   \label{profile2}
\end{eqnarray}
where $\delta N$ is a free fitting parameter.
When $\delta N=0$ Eq.\ (\ref{profile2}) is simply the discrete
version of Eq.\ (\ref{profile1}) (note that now the system is defined for $n=1,\dots,N$). 

In the lack of a consistent scheme to calculate corrections, the above amendment to the fitting
formula is not rigorous. Justification can stem from a direct comparison with
exact results. In the following we first discuss in detail the exactly solvable $S=1/2$
chain as a benchmark case. We demonstrate that the Ansatz in Eq.\ (\ref{profile2}) is
capable of reproducing the (numerically) exact $S=1/2$ critical exponents to high precision
even with $\delta N=0$. When $\delta N$ is also fitter for, the accuracy improves by an order of
magnitude.

\section{The $S=1/2$ chain: Bethe Ansatz results}

The $S=1/2$ Heisenberg chain in a magnetic field can be solved exactly by the Bethe
Ansatz (BA) method.\cite{Grif-Yan-Yan} The BA itself cannot yield the correlation functions
and critical exponents directly, but assuming conformal invariance the operator content can be read off from
the structure of the low-energy excitations above the ground state.\cite{Cardy} These latter
can be determined by a systematic calculation of the finite size corrections to the
$L=\infty$ BA equations (assuming periodic boundary condition).\cite{Woy-Eck-deVega}
The energy and momentum of the lowest energy (primary) states (with respect to those of the
ground state, $E_g$ and $P_g$) can be cast into the form
\begin{eqnarray}
   \delta E&=& E_{\bf a}- E_g =
   \frac{2\pi v}{N}(\Delta^+ +\Delta^-),
   \label{E_diffhalf} \\
   \delta P&=& P_{\bf a}-P_g = Q +\frac{2\pi}{N}(\Delta^+ -\Delta^-) ,
   \label{P_diffhalf}
\end{eqnarray}
where ${\bf a}=\{d,l\}$ is a shorthand for two integer topological quantum
numbers labeling the states. The ${\cal O}(1)$ momentum term is
\begin{eqnarray}
   Q = Q_{\bf a}=2k_F d +\pi l,
   \label{Qhalf}
\end{eqnarray}
with $k_F$ defined in Eq.\ (\ref{m_kF}). The "conformal dimensions" $\Delta^\pm$
read
\begin{equation}
   \Delta^\pm = {1\over 2} \left[Zd \pm \frac{l}{2Z} \right]^2,
   \label{Deltashalf}
\end{equation}
where $Z$ is the dressed charge. The topological quantum numbers have a direct physical
interpretation in the LL representation: $l$ ($d$) denotes the number of fermions
added to (transferred from the left Fermi point to the right in) the band.

\begin{figure}[tb]
\includegraphics[scale=.45]{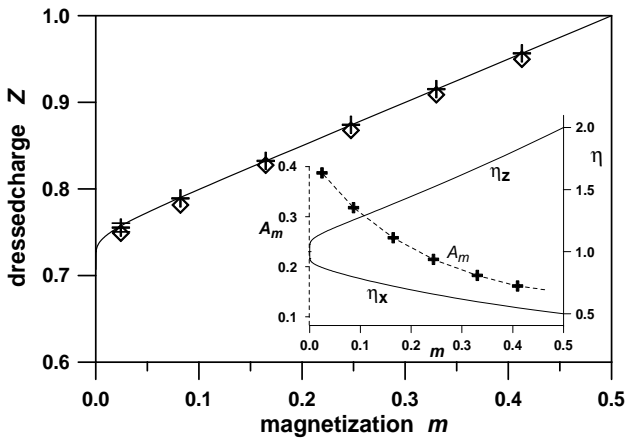} 
\caption{The dressed charge as a function of the magnetization in an $S=1/2$ chain;
$\diamond$ denotes results using the naive fit
$\delta N=0$, $+$ denotes fits where $\delta N$ is optimized.
Solid line is the Bethe Ansatz result. Inset shows the critical
exponents and the numerically determined amplitude $A_m$. At $m=0$ the dressed
charge is $Z=1/\protect\sqrt{2}$.} 
\label{fig:zhalf} 
\end{figure}

The critical exponents appearing in the one- and two-point correlation functions
can be expressed with the conformal dimensions $\Delta^\pm$. In general a physical
operator such as $S^z$ decomposes into all operators which are not forbidden
by conservation (selection) rules. In particular, for $S^z$ the number of fermions
in the band is conserved, thus necessarily $l=0$. The asymptotic decay is determined
by the operator which has the lowest critical exponent, i.e., $d=1$ giving
\begin{equation}
   \eta_z =\left. 2(\Delta^+ +\Delta^-) \right|_{l=0,d=1} =2 Z^2,
   \label{etaz}
\end{equation}
and an oscillation $2k_F$ by Eq.\ (\ref{Qhalf}).
The transverse critical exponents $\eta_x$ is determined by the operator
$l=1, d=0$, i.e.,
\begin{equation}
   \eta_x =\left. 2(\Delta^+ +\Delta^-) \right|_{l=1,d=0} =\frac{1}{2 Z^2},
   \label{etax}
\end{equation}
with the associated oscillation frequency $Q=\pi$. Equations (\ref{etaz}) and
(\ref{etax}) imply $\eta_z \eta_x=1$.

In the BA approach the dressed charge $Z=Z(B)$ is determined by a set of
integral equations for the density of rapidities $\rho(x)$, and the dressed charge
function $Z(x)$,\cite{Par-Schm-Cab}
\begin{eqnarray}
    \rho(x) &=& g(x)+\int_{-B}^B K(x-x') \rho(x') dx',
       \label{Rhalf} \\
    Z(x) &=& 1+\int_{-B}^B K(x-x') Z(x') dx',
       \label{Zhalf}
\end{eqnarray}
where
\begin{equation}
   g(z)=\frac{1}{\pi(1+z^2)}, \qquad K(z)=\frac{-2}{\pi(4+z^2)}.
\end{equation}
The limits of integration are defined implicitly through the constraint
\begin{equation}
   \int_{-B}^B \rho(x') dx' = 1/2-m.
   \label{Bhalf}
\end{equation}

The coupled integral equations in Eqs.\ (\ref{Rhalf}), (\ref{Zhalf}) and (\ref{Bhalf})
are to be solved numerically.\cite{Par-Schm-Cab} First, assuming $B$ is given, Eq.\ (\ref{Rhalf})
is solved for $\rho$. This is inserted into Eq.\ (\ref{Bhalf}) to find the associated
value of $m$. When $m$ is given, as in our case, this procedure can be iterated
to find $B$ as a function of $m$ at arbitrary precision.
Finally Eq.\ (\ref{Bhalf}) is solved for the function $Z(x)$ and the dressed charge
$Z=Z(B)$ is determined. The numerically determined dressed charge as a function of
the magnetization is plotted in Fig.\ \ref{fig:zhalf}. In the next section this will serve as
a reference curve to check the accuracy of the fitting Ansatz in Eq.\ (\ref{profile2}).

\section{The $S=1/2$ chain: DMRG results}

\begin{figure}[tb]
\includegraphics[scale=.45]{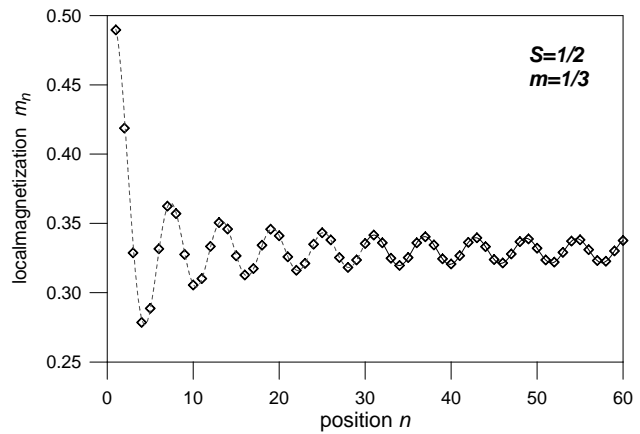} 
\caption{Magnetization profiles determined by the DMRG for the $S=1/2$ chain with 
$N=120$, $M=160$ and magnetization $m=S^z_{\rm tot}/N\approx 1/3$. Only half of the
chain shown - the other half is mirror symmetric.} 
\label{fig-pa} 
\end{figure}

In order to test the fitting formula Eq.\ (\ref{profile2}) we calculated the
magnetization profile in the ground state of finite chain segments with open
boundary condition using the density matrix renormalization group method.\cite{Whi}
The chain length was $N=120$ and we kept $M=160$ states. The truncation
error was found to be in the range $1-P_M=10^{-13}-10^{-11}$. We applied the finite
lattice algorithm with four iteration cycles which was found sufficient for convergence.
An example of the magnetization profile as determined by the DMRG is depicted in Fig.
\ref{fig-pa}.

\begin{figure}[tb]
\includegraphics[scale=.45]{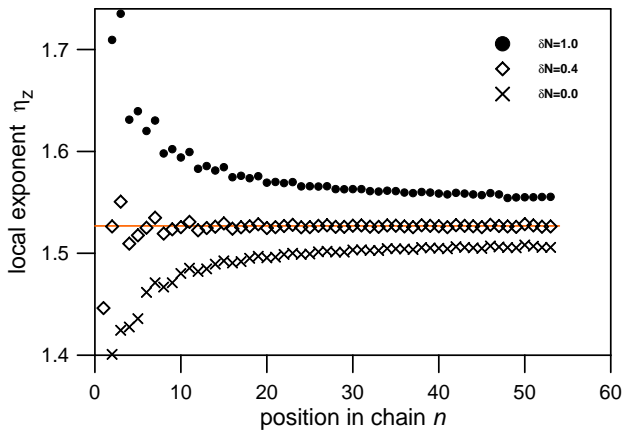} 
\caption{Stability of local fit parameters as a function of local position in the chain
for $S=1/2$, $L=120$, $m\approx 1/4$. Local fit parameters at position $n$ are defined
by fitting for sites $i=n,\dots,n+3\lambda$, with $\lambda=2\pi/k_F=1/(S-m)$ the wavelength
of the oscillations. The highest stability of the parameters is achieved for
$\delta N\approx 0.4$. Equation (\protect\ref{m_kF})
is satisfied up to $2.10^{-4}$.} 
\label{fig:fitting} 
\end{figure}

It is worth discussing the fitting procedure itself in some detail. The Ansatz in
Eq.\ (\ref{profile2}) has five parameters: $m,A_m,k_F,\delta N$ and $\eta_z$. Although
$S^z_{\rm tot}$ is a conserved quantity, and thus can be set to a given value
in the DMRG, due to the open boundary condition the control over the exact (bulk) value
of $m$ is lost. While
it is true that for long enough chains $m$ will be close to $S^z_{\rm tot}/N$, the
finite size deviation should be tracked during the fitting procedure. Similarly, although
$k_F$ is a well-specified function of $m$ in the bulk [see Eq.\ (\ref{m_kF})], it was found advantageous
to keep it as an independent fit parameter, and only use Eq.\
(\ref{m_kF}) a posteriori as a consistency check. On the other hand, 
it is better to avoid fitting on $\delta N$ directly. Instead the best working alternative seems
to be making a four-parameter fit on $m,A_m,k_F$ and $\eta_z$, while $\delta N$ kept fixed.
The optimal value of $\delta N$ is the one which yields the highest stability with respect to
local fits, i.e., calculating the four fitting parameters from a small number of sites
at different locations in the chain. An example of this procedure is shown in Fig.\
\ref{fig:fitting}. We found that the optimal value of $\delta N$ is a weak function of $m$,
being about $\delta N\approx 0.5$ for $m=0$, and decreasing monotonically to
$\delta N\approx 0.4$ for $m=1/2$. This is more or less consistent with the value $\delta N=1/2$
used in the bosonization approach of Ref.\ \onlinecite{Hik-Fur}.

Having obtained $\eta_z$ using the above fitting procedure, the fundamental quantity
of the theory, the dressed charge $Z$, can be calculated from Eq.\ (\ref{etaz}).
Figure \ref{fig:zhalf} shows the numerically determined dressed charge as a function
of $m$. The relative error of the fitting procedure is under $0.1\%$, except for
very small $m$ values where logarithmic corrections to the fitting formula and $Z$ are
expected, and the fitting procedure loses stability (see the error bar in the figure
at $m=0.03$).
Otherwise, the agreement with the exact values is very good and stable for
$0.03\alt m<1/2$. For
comparison Fig.\ \ref{fig:zhalf} also shows the estimate of $Z$ when the naive
scattering length $\delta N=0$ is used. Even with this the error is within $1\%$, except for
very small $m$.

We conclude that for $S=1/2$ the LL Ansatz for the
magnetization profile, Eq.\ (\ref{profile2}), is a very efficient tool in calculating
the critical exponents (dressed charge). The final result is highly
accurate already with $\delta N=0$, but an additional increase in precision can be achieved
by adding the scattering length as an extra fit parameter.

\section{The $S=1$ chain}

Encouraged by the success of the fitting procedure for $S=1/2$, in this section we
apply it to the $S=1$ Heisenberg chain. The forthcoming analysis is not intended to
prove that the $S=1$ chain in its magnetized regime is a Luttinger liquid - this
has been done convincingly already.\cite{Tsvelik,Chi-Gia,Kon-Fen,Sak-Tak,Campos-Venuti} Instead,
our starting point is the assumption
that we have a one-component Luttinger liquid, and then use LL theory and the
numerically
calculated magnetization profiles to determine the value of $Z$ and the critical
exponents with high accuracy.

Although the spin-1 problem cannot be treated rigorously, there are two limits where
clear theoretical predictions exist. In the high magnetization limit near saturation
the physics can be understood by regarding the system as a dilute gas of magnons
created in the ferromagnetic vacuum.\cite{Sak-Tak} Magnons behave as bosons with short-range
repulsive interaction. In the dilute limit the exact form of the interaction is
irrelevant and an essentially hard-core boson description becomes valid. In 1D
hard-core bosons are equivalent to free spinless fermions, which imply an LL
parameter $Z=1$ with correlation exponents $\eta_z=2$ and $\eta_x=1/2$ as $m\to 1$.

There is a similar argument in the $m\to 0$ limit. Below the critical magnetic field
$h_c<\Delta_g$, where $\Delta_g$ is the Haldane gap, the elementary excitations are massive
spin-1 bosons.\cite{Aff-Sor} At $h_c$ these bosons condensate, but due to the interboson
repulsion the
magnetization only increases gradually. Near $m\agt 0$ the boson density is low and
by the above argument we again obtain $Z=1$, $\eta_z=2$ and $\eta_x=1/2$. Note that
this argument only works for $S=1$ with a Haldane gap. For $S=1/2$ for which
the nonmagnetized ground state is massless, the elementary excitations form a strongly
interacting dense gas with $Z=1/\sqrt{2}$ as is given by the Bethe Ansatz, and the
SU(2) symmetry at $m=0$. Alternative theories for $S=1$, based on either a Majorana fermion
representation\cite{Tsvelik} or on the bosonization of spin-1/2 ladders\cite{Chi-Gia}
lead to a similar conclusion in this limit.

Although the magnetized $S=1$ chain is a Luttinger liquid, this classification only
applies at low energies and long distances. Indeed, at higher energies and shorter
distances the $S=1$ chain produces features which cannot be understood within the framework of
LL theory. These features, absent in $S=1/2$ chains, stem from
the additional degrees of freedom staying massive for $S>1/2$. These degrees of
freedom have signature both in the energy spectrum and correlations, and manifest
themselves in the numerical, finite chain calculations. Their origin can be easily
understood in the low-magnetization limit.
At zero magnetic field the system possesses
an energy gap, the Haldane gap. The lowest excited states form a triplet branch with
a minimum energy at momentum $k=\pi$. The operator
$S_i^z$ has large matrix elements between the ground state and the $S^z_{\rm tot}=0$
component of this triplet. This leads to an exponentially decaying alternating
(antiferromagnetic) behavior in the longitudinal correlation function. When the
magnetic field is switched on, the Zeeman energy splits the triplet branch, and
at $h=\Delta_g$ the $S^z_{\rm tot}=1$ component at $k=\pi$ crosses over with the ground
state. However the $S^z_{\rm tot}=0$ component remains in the spectrum (at energy
$\approx\Delta_g$) and still contributes to the short-range longitudinal correlation
functions. As a consequence the two-point function shows a crossover from a seemingly
exponential decay on short distances caused by the massive mode to an algebraic decay
determined by the soft, LL mode on longer distances.

\begin{figure}[tb]
\includegraphics[scale=.46]{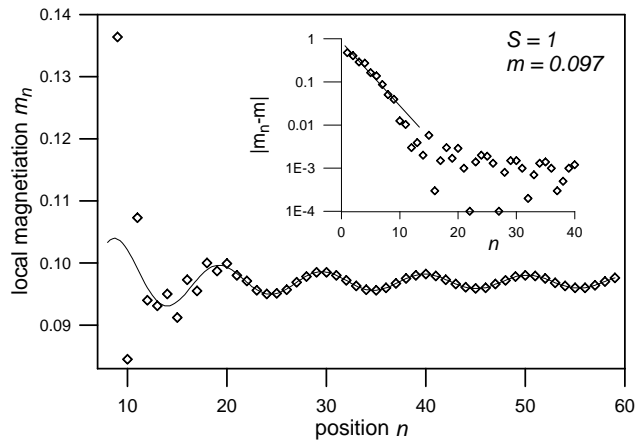} 
\caption{Magnetization profiles determined by the DMRG for the $S=1$ chain with 
$N=120$, $M=160$ at the bulk magnetization $m\approx 0.097$. Only half of the
chain shown - the other half is mirror symmetric. The LL oscillation is clearly
visible above $n > 20$. Solid line is the LL fit. Inset shows the end of the chain
on a log-lin scale featuring
the exponentially decaying initial oscillation which crosses over to LL behavior at
$n\approx 12$.} 
\label{fig-pb} 
\end{figure}

There is a similar
effect in the one-point function we consider here. Near the chain's end there is an exponentially
localized effective $S=1/2$ degree of freedom, the so-called "end spin", which also survives in the
magnetized regime, at least when the magnetization is not too high.\cite{Yam-Miy} Its presence
produces a crossover from exponential decay to algebraic decay in the one-point
correlation function as is illustrated in Fig.\ \ref{fig-pb}. As the magnetization
increases the massive modes rise in energy, and have a less and less significant
impact on the low-energy physics. At the same time the end spins gradually dissolve
and disappear in the bulk as was observed by Yamamoto and Miyashita.\cite{Yam-Miy}

In order to measure the critical exponent $\eta_z$ we used the DMRG algorithm
with $N=120$ and $M=160$ with five iteration cycles. The truncation error varied in
the range $1-P_m=10^{-12}-10^{-8}$, the calculation being more precise in the high
magnetization regime. A limited number of runs with $N=240$, $M=300$
was done to check the numerical precision and to obtain results at points where longer
systems were needed. We applied the fitting procedure described above
and care was taken to stay in the bulk of the chain sufficiently far from the ends.
Since for $m=0$ the correlation (localization) length of the end spin is about
6 lattice sites, which becomes even shorter for $m>0$, a chain length $N=120$
was found sufficient.

We observed that the fitting procedure is somewhat less accurate here than for $S=1/2$,
meaning that finite size corrections to the Luttinger liquid profile Eq.\
(\ref{profile2}) are more important in for $S=1$. The fitting procedure becomes especially
unreliable below $m\approx 0.05$ and around $m\approx 0.5$. In the former case
the increasing wavelength, which becomes comparable to the system size, whereas
in the latter case the vanishing prefactor $\cos[2k_F(n-(N+1)/2)]\to 0$
can be blamed for the numerical difficulty. Far enough from these problematic regions the
relative error of the calculated exponent was estimated to be less than $1\%$.

Beyond measuring the local magnetization profile $m_n$, for $S=1$ there is another,
independent quantity whose profile can also be measured easily. This is the local
quadrupole moment $q_n=\langle (S^z_n)^2 \rangle$. Note that for $S=1/2$ this
quantity is redundant and thus carries no
additional information. For $S=1$, however, the quadrupole profile provides
us an alternative way to measure the critical exponent which in many cases became
even more precise than the one obtained through the magnetization profile. The quadrupole
profile is expected to behave according to the same scaling form Eq.\ (\ref{profile2})
with the replacement $m\to q$, $A_m\to A_q$. 

\begin{figure}[tb]
\includegraphics[scale=.48]{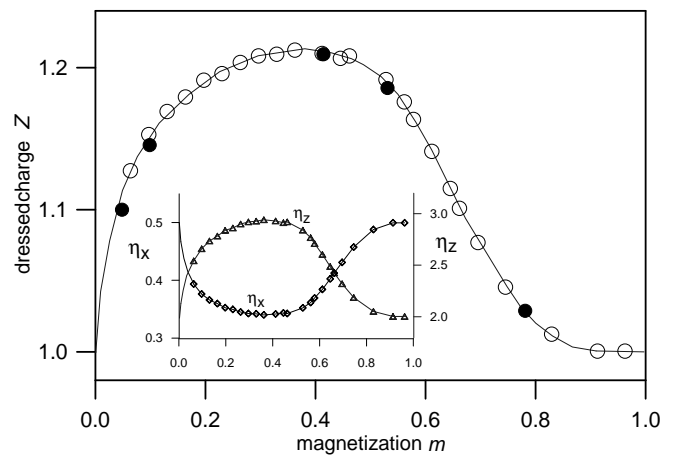} 
\caption{Dressed charge $Z$ determined numerically for the $S=1$ chain.
Open symbols denote results for $L=120$, closed symbols for $L=240$.
The estimated error is comparable to the symbol size. Solid line is only a
guide to the eye. Inset shows
the longitudinal and transverse critical exponents $\eta_z$ and $\eta_x$, resp.} 
\label{fig:expon} 
\end{figure}

Figure \ref{fig:expon} shows the dressed charge determined using Eq.\ (\ref{etaz})
from the measured $\eta_z$ exponent of magnetization and quadrupole profiles.
The inset also shows the corresponding critical
exponents $\eta_z$ and $\eta_x$ as a function of the bulk magnetization. We see that
for $m\to 1$ the predicted value $Z=1$ is reached very rapidly. We analyzed the
$m$-dependence close to $m=1$ and found it to be describable with a power law with a
rather large exponent around 4, although a scaling faster than any power law cannot be
excluded either. Due to the lack of sufficient numerical precision near $m=1$ we were
unable to resolve this question reliably. 

For $m\to 0$ the conclusion is also somewhat vague because the $m$-dependence is
very steep and
due to numerical difficulties we were unable to approach this limit closely enough.
However, a $Z=1$ value predicted by the theory seems highly consistent with our data.
Assuming this, our
numerical values seem to indicate an $m$-dependence $Z-1\sim m^\alpha$, with
$\alpha=0.5\pm 0.1$. Knowing that $m$ scales above the gap as
$m\sim \sqrt{h-\Delta_g}$,\cite{Sak-Tak-98}
this would imply $Z-1\sim (h-\Delta_g)^\beta$ for small $h$ with $\beta\approx 1/4$.

Between the two limits
$Z$ is larger, reaching its largest value $Z\approx 1.21$ at $m\approx 0.36$. This
behavior should be compared to that of $S=1/2$, where $Z<1$ in the whole regime (see
Fig.\ \ref{fig:zhalf}). Earlier data on the critical exponents available in the literature,
which were determined by other methods such as the finite-size scaling analysis of the energy
spectrum\cite{Sak-Tak,Usa-Sug} or the staggered structure factor,\cite{Campos-Venuti} are consistent
with our results.

\begin{figure}[tb]
\includegraphics[scale=.45]{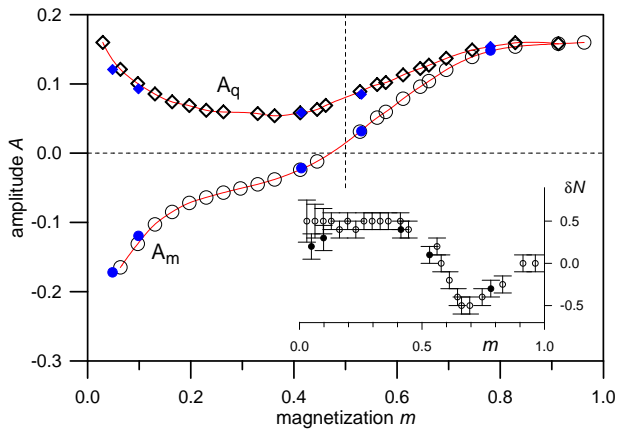} 
\caption{Magnetization and quadrupole moment amplitudes for the $S=1$ chain. Open symbols
denote results for $L=120$, closed symbols for $L=240$. Inset shows the best estimate
of $\delta N$.} 
\label{fig:ampli} 
\end{figure}

The numerically determined critical amplitudes are depicted in Fig.\ \ref{fig:ampli}.
The amplitude of the magnetic moment fluctuations $A_m$ increases monotonically as a
function of $m$. However, the rate of increase is not smooth as it is seen in the figure.
$A_m$ changes sign somewhere close to $m=1/2$. Where $A_m\approx 0$
the observable fluctuations are governed by the next smallest critical exponent, and thus
a precise measurement is beyond our method.
In contrast with this, the amplitude of the quadrupole fluctuations $A_q$ remains positive
in the whole magnetized regime. We observe that it decreases for small $m$,
reaching its minimal (still positive) value at $m\approx 0.36$. Above this it increases
and saturates for $m\to 1$.

Finally, it is interesting to note that the optimal value of the scattering length $\delta N$
changes considerably as $m$ varies. As the inset of Fig.\ \ref{fig:ampli} shows that $\delta N$
is around $0.5$ for small $m$, then decreasing to $\delta N\approx -0.5$ at
$m\approx 0.7$, then increasing again to $\delta N\approx 0$ at $m=1$. There is a relatively large uncertainty
in the optimal value determined, but this imprecision has little impact on the estimated
value of $Z$ and the correlation exponents.

\section{Summary}

In this paper we have analyzed the critical fluctuations in open $S=1/2$ and $S=1$ Heisenberg
chains in their magnetized regime. The low energy physics is a one-component Luttinger liquid in
both cases. We have calculated the LL's characteristic dressed charge
and the amplitude of the leading critical fluctuations. Our method consisted of
determining numerically the magnetization and quadrupole operator profiles and
applying a fitting procedure based
on conformal invariance. The method has been thoroughly tested on the Bethe Ansatz solvable
$S=1/2$ chain, confirming its reliability and high precision. In the $S=1$ case,
where the exact solution is unknown the method provided high-precision estimates
of the critical exponents, justifying and complementing earlier results derived by
alternative methods. We also determined critical amplitudes which has much less been
studied so far for these systems.

Beyond calculating the characteristic parameters with high accuracy our results also
allow us to make a detailed comparison between the $S=1/2$ and $S=1$ chains. Although
both are Luttinger liquids, the respective dressed charges as functions of the
magnetization differ quite considerably. The only limit where the two systems become
equivalent is the full saturation limit, $m\to S$, where $Z\to 1$. Otherwise, for $S=1/2$
the dressed charge is $Z<1$, while for $S=1$ it is $Z>1$. There are interesting
differences in the behavior of the critical amplitudes and the scattering lengths at
the chain ends, as well.

We thank F.\ Igl\'oi and J. S\'olyom for valuable discussions. This work was partially
supported by the Hungarian Scientific Research Found (OTKA) under grant Nos. T30173,
T43330 and F31949, and by the Bolyai Research Scholarship Program. The numerical
calculations were done at the NIIF Supercomputing Facility.

\end{document}